\documentclass[11 pt]{article}
\usepackage{ucs} 
\usepackage[utf8x]{inputenc} 

\title{...}
\title{\LaTeX}  
\date{}  
\author{}

\usepackage{amsmath, amsfonts, amssymb}
\usepackage{graphicx}
\usepackage{indentfirst}
\usepackage{threeparttable}
\usepackage{url}

\def\la{\;
\raise0.3ex\hbox{$<$\kern-0.75em\raise-1.1ex\hbox{$\sim$}}\; }
\def\ga{\;
\raise0.3ex\hbox{$>$\kern-0.75em\raise-1.1ex\hbox{$\sim$}}\; }

\newcommand{\kms}{km\,s$^{-1}$}
\newcommand{\ms}{m\,s$^{-1}$}

\newcommand{\dmm}{$\Delta\mu/\mu$}

\begin{document}

\title{\LARGE \bf  Indications of electron-to-proton mass ratio
 variations in the Galaxy. II.
3~mm methanol lines toward Sgr\,B2(N) and (M).
}

\maketitle

\begin{center}
J. S. Vorotyntseva$^*$\footnote{j.s.vorotyntseva@mail.ioffe.ru}, S. A. Levshakov$^*$, C. Henkel$^{**}$\\
$^*$Ioffe Institute, St. Petersburg, 194021 Russia\\
$^{**}$Max Planck Institut f\"ur Radioastronomie (MPIfR), Auf dem 
H\"ugel, D-53121 Bonn, Germany
\end{center}

\bigskip\noindent

\begin {abstract} 
Differential measurements of the fundamental constant  
$\mu = m_{\rm e}/m_{\rm p}$ 
(the electron-to-proton mass ratio)
for two sources near the Galactic Center~-- the Sgr\,B2(N) and B2(M) molecular
clouds~-- 
suggest that $\mu$  is lower in these clouds than its laboratory value.
Based on observations of methanol (CH$_3$OH) emission lines in the 80--112 GHz range (data from the IRAM 30-m  telescope), a weighted mean value 
$\langle \Delta \mu/\mu \rangle \equiv 
\langle (\mu_{\rm obs} - \mu_{\rm lab})/\mu_{\rm lab}\rangle =
(-2.1\pm0.6)\times10^{-7}$ ($1\sigma$)  was obtained for Sgr\,B2(N)
at the sample size $n = 9$.
This value of $\Delta\mu/\mu$ 
has the same sign as the result of recent measurements of methanol lines in the higher frequency range of 542--543 GHz 
(data from the {\it Herschel} space telescope) for Sgr\,B2(N):
$\langle \Delta\mu/\mu \rangle = (-4.2\pm0.7)\times10^{-7}$ 
(sample size $n = 2$).
\end{abstract}

\section{Introduction}
\label{Sec1}

Fundamental physical constants are naturally included in all laws of physics; their numerical values ​​should reflect the properties of the world around us,
but their origin remains unexplained up to now.
In 1937, Dirac \cite{Dir37} suggested that
in a dynamically evolving Universe, physical constants could also be dynamical
variables corresponding to the local age of the Universe.
However, numerous laboratory experiments with atomic, and more recently nuclear, clocks have not revealed any changes. 
The most stringent upper limit currently available, for example, for the fine structure constant $\alpha = e^2/\hbar c$, is set at 
$\Delta\alpha/\alpha < 10^{-19}$~yr$^{-1}$ ($1\sigma$)\footnote{All estimates of physical quantities are given at
the $1\sigma$ confidence level.}.

Other models consider the dependence of the physical constants on the values ​​of the local density
of baryonic matter \cite{Kh04} or on the strength of the gravitational field \cite{Flam08}.
However, in order to obtain a self-consistent picture with quantum physics, the variations of the constants
cannot be arbitrary \cite{Ub18}.
For example, satisfying the requirement of conservation of energy entails the inclusion of additional 
scalar fields into the theory,
which may be associated with another, unknown (dark) form of matter \cite{Bar02}.
  If such a coupling does exist, then one might expect that the values ​​of some physical constants
will be functionally related to local variations in the density of dark matter.
This picture is typical, for example, for different regions of the Galaxy~-- the center and the periphery.
A detailed analysis of available measurements and theoretical models in this field is given in a recent review \cite{Uz24}.

Astronomical observations allow us to significantly expand laboratory experiments
and probe dimensionless constants under various environmental conditions, locally and on cosmological time scales.
It turned out that in this case the most sensitive quantity
to hypothetical spatial and temporal changes in the constants is  the electron-to-proton mass ratio, $\mu = m_{\rm e}/m_{\rm p}$.
The tightest cosmological constraint 
on a time scale of $\sim 10$ billion years, 
based on an analysis of transitions in methanol (CH$_3$OH), yields 
$\Delta\mu/\mu \equiv (\mu_{\rm obs} - \mu_{\rm lab})/\mu_{\rm lab} < 5\times10^{-8}$ \cite{Kan}.
An upper limit of the same order of magnitude ($\sim 10^{-8}$)
was obtained from observations of the CH$_3$OH and $^{13}$CH$_3$OH lines in the disk of our Galaxy at relatively large galactocentric distances,
$D_{\rm GC} \sim 8$ kpc \cite{L11}-\cite{VL24}.

However, the vicinity of the central part of the Galaxy has not been studied in detail until recently.
Two attempts have been made to test
one part of the Einstein equivalence principle~--
the local position invariance~-- in the vicinity of the supermassive black hole (SMBH) Sgr\,A$^\ast$ using atomic absorption lines in the spectra of stars located in the vicinity of the SMBH \cite{A19, Hees20}.
Both attempts were aimed at estimating variations in the fine structure constant $\alpha$ at a sensitivity level of $\Delta\alpha/\alpha \la 10^{-5}$.
A two-order-of-magnitude higher sensitivity to variations in fundamental constants can be achieved in radio astronomical spectral observations, in particular in estimating $\Delta\mu/\mu$ from microwave spectra of various molecules. Such assessments have not yet been carried out in the central regions of the Galaxy.
In this regard, an attempt was made to fill this gap in \cite{VL25}, where variations of $\Delta\mu/\mu$ were estimated using methanol lines from the high-frequency range
$\Delta f = 542-543$~GHz  observed with the {\it Herschel} space 
telescope\footnote{ {\it Herschel} is an ESA space observatory, 
equipped with instruments developed by leading European centers with significant participation from NASA.}
toward the molecular cloud Sagittarius (Sgr) B2(N), which is located
near the Galactic Center (SMBH), at a distance of 
$D_{\rm GC} \sim 130$~pc  \cite{RMZ}.
The result of this preliminary analysis of the relative positions of the three methanol lines
showed a possible decrease in $\mu$ at a statistically significant level:
$\Delta\mu/\mu = (-4.2\pm0.7)\times10^{-7}$  \cite{VL25}. 
 
In the present paper, we continue these studies using a different set of methanol lines, a different frequency range, a different telescope, a different spectral
 resolution,
and a new object, Sgr\,B2(M), along with Sgr\,B2(N).

\begin{table*}[h!]
\centering
\label{T1}
\caption{Parameters of the selected CH$_3$OH torsion-rotation  transitions.
The measured $f_{lab}$ and calculated $f_{cal}$
line frequencies (in MHz) and  the low level energies $E_\ell$ are taken from the indicated references.
Given in parentheses are the $1\sigma$ errors in the last digits.
}
\begin{tabular}{c l  r@.l r@.l r@.l r@.l}
\hline
\multicolumn{1}{c}{\textrm{No.}} &
\multicolumn{1}{c}{\textrm{Transition}} &
\multicolumn{2}{c}{\textrm{$E_\ell$,}} &
\multicolumn{2}{l}{\textrm{Measured}} &
\multicolumn{2}{l}{\textrm{Calculated}} &
\multicolumn{2}{c}{\textrm{$Q$}} \\
 & \multicolumn{1}{c}{\textrm{ $J_{K_u} \to J_{K_\ell}$ }}  & 
 \multicolumn{2}{c}{\textrm{ cm$^{-1}$ }} &
 \multicolumn{2}{l}{\textrm{frequencies, $f_{lab}$ } } &
\multicolumn{2}{l}{\textrm{frequencies, $f_{cal}$ } }   \\
\hline
\multicolumn{10}{c}{\textit{ First subband $80.0-89.9$ GHz } } \\
1&$7_{2} \to 8_{1}A^-$&68&7$^a$&80993&160(50)$^a$&80993&245(12)$^c$  &
4&0$^d$ \\ 
2&$13_{-3} \to 14_{-2}E$&187&6$^a$&84423&810(50)$^a$& 84423&769(12)$^c$ &
5&8$^e$ \\
3&$7_{2} \to 6_{3}A^-$&68&5$^a$&86615&600(5)$^b$&86615&574(13)$^c$ &
7&3$^d$ \\
4&$7_{2} \to 6_{3}A^+$&68&5$^a$&86902&949(5)$^b$&86902&916(5)$^c$  &
7&3$^d$\\
5&$15_{3} \to 14_{4}A^+$&225&2$^a$&88594&960(50)$^a$&88594&787(11)$^c$ &
2&3$^e$ \\
6&$15_{3} \to 14_{4}A^-$&225&2$^a$&88940&090(50)$^a$&88939&971(11)$^c$ &
2&3$^e$ \\[4pt]
\multicolumn{10}{c}{\textit{ Second subband $89.9-100.2$ GHz } } \\
7&$8_{3} \to 9_{2}E$&88&1$^a$&94541&787(50)$^a$&94541&785(14)$^c$ &
0&2$^d$ \\
8&$8_{0} \to 7_{1}A^+$&54&9$^a$&95169&463(10)$^b$&95169&391(11)$^c$ &
$-1$&9$^d$ \\
9&$2_{1} \to 1_{1}A^-$&11&7$^a$&97582&804(7)$^b$&97582&798(2)$^c$ &
1&0$^d$ \\[4pt]
\multicolumn{10}{c}{\textit{Third subband $100.2-114.6$ GHz } } \\
10&$13_{2} \to 12_{3}E$&159&0$^a$&100638&900(50)$^a$&100638&872(15)$^c$ &
0&9$^e$ \\
11&$17_{-2} \to 17_{1}E$&261&4$^a$&111626&530(50)$^a$& 111626&514(15)$^c$ & $-1$&8$^e$ \\
\hline
\multicolumn{10}{l}{\footnotesize References: $a\, -$ \cite{Xu1997}; $b\, -$ \cite{Mull}; 
$c\, -$ \cite{Xu2008}; $d\, -$ \cite{J11}; $e - Q$ is calculated in the present 
work.  } \\
\end{tabular}
\end{table*}

\section{Targets and observations}
\label{Sec2}

The molecular cloud Sgr\,B2 is the largest molecular cloud in the central part of
the Galaxy with a mass of $M_{\rm Sgr B2} \sim 10^7 M_\odot$ \cite{Schw},
 as well as one of the most massive clouds in the Galaxy as a whole \cite{Sc}.
This is an active star formation region consisting mainly of four extensive complexes~--
Sgr\,B2(N) in the northern part, Sgr\,B2(M) in the middle part, Sgr\,B2(S) in the southern part,
and possibly the G+0.693-0.027 region located slightly to the east of Sgr\,B2(N). 

In this work, we use the publicly
available Sgr\,B2(N) and (M) spectra, which were obtained using the IRAM 30-m telescope\footnote{The Institute for Radio Astronomy in the
Millimeter Range (IRAM) is an international research institute and Europe's leading center for radio astronomy.} in 2004-2005 \cite{Bel}.
Detailed data on these observations are given in the original work \cite{Bel},
and the spectra are available at \cite{Bel13http}.
Below we will list only the main parameters of these observations.

The frequency range of 80--116 GHz (atmospheric window of 3 mm) was divided into four subbands (80.0--89.9 GHz, 89.9--100.2 GHz, 100.2--114.6 GHz and 114.6--116.0 GHz); observations were carried out simultaneously using two heterodyne receivers connected to the VESPA spectrometer-autocorrelator (VErsatile Spectrometric and Polarimetric Array).
The spectral resolution was 312.5 kHz, the bandwidth~-- 420 MHz, the aperture $HPBW\sim 30^{\prime\prime}$, and the pointing accuracy of the telescope $2^{\prime\prime} - 3^{\prime\prime}$ (rms).

Coordinates of objects~-- 
$R.A. = $17$^{\rm h}$47$^{\rm m}$$20.\!^{\rm s}0$, 
$Dec. = -28^{\circ}$22$^{\prime}$$19.\!^{\prime\prime}0$ (J2000),
radial velocity of the cloud $V_{\rm LSR} = 64$~\kms\ for Sgr\,B2(N);
and $R.A. = $17$^{\rm h}$47$^{\rm m}$$20.\!^{\rm s}4$, 
$Dec. = -28^{\circ}$23$^{\prime}$$07.\!^{\prime\prime}0$ (J2000), 
$V_{\rm LSR} = 62$~\kms\ for Sgr\,B2(M).

It should be noted that the central coordinates and the aperture of the telescope in the {\it Herschel} observations of Sgr\,B2(N)  used in \cite{VL25} are almost identical to the parameters of the IRAM: 
$R.A. = $17$^{\rm h}$47$^{\rm m}$$20.\!^{\rm s}06$,
$Dec. = -28^{\circ}$22$^{\prime}$$18.\!^{\prime\prime}33$ (J2000),
$HPBW  \sim 40^{\prime\prime}$.
At the same time, the spectral resolution in the high-frequency range at 0.6~mm
({\it Herschel}) was 3 times higher than at 3~mm (IRAM).

The obtained spectra were calibrated in antenna temperatures ($T_{\rm mb}$)
and frequencies $f_{obs}$ using 
the standard software package CLASS\footnote{Continuum and Line Analysis Single-dish Software \cite{CLASS}.}.

\begin{figure}[h!]
\vspace{-2cm}
\centering
\includegraphics[width=0.9\textwidth]{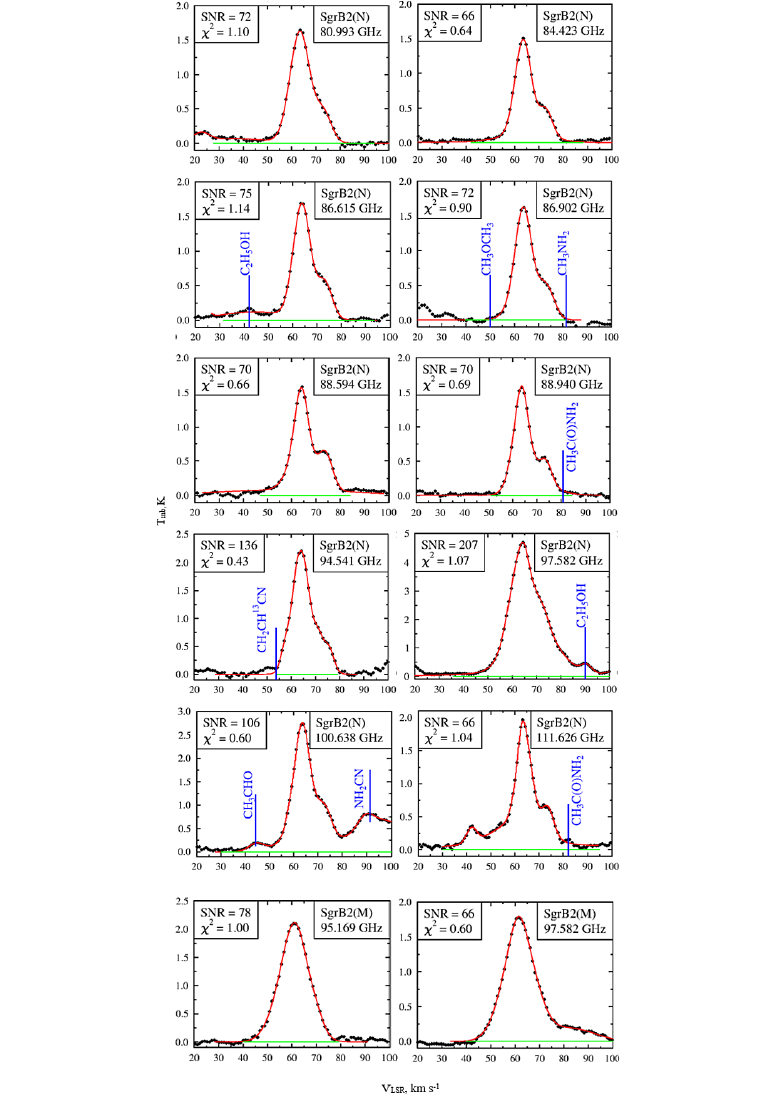}
\caption{\footnotesize
Selected methanol (CH$_3$OH) lines toward Sgr\,B2(N) and (M) obtained
with the IRAM 30-m telescope in a main-beam temperature scale  \cite{Bel}.
The original spectra are displayed by black dots, whereas the fitting curves
are shown in red. The horizontal green lines indicate the ranges in which $\chi^2$ was calculated. The vertical blue bars mark the identified blends in
accord with Figs.~2 and 5 from  \cite{Bel}. A weaker methanol emission from
the second hot core N2 (offset by $6^{\prime\prime}$) is seen at about 
 75~\kms\ in the Sgr\,B2(N) spectra. 
}
\label{F1}
\end{figure}

\begin{table*}[h!]
\centering
\label{T2}
\caption{Fitting parameters of the selected methanol lines~-- the central
frequency $f_{obs}$ and the full width at half maximum $FWHM$~-- from
the range 80--112 GHz in spectra of  Sgr\,B2(N) and (M).
The lines are numbered and grouped according to the observable frequency
subbands as in Table~\ref{T1}.
$As$ is the asymmetry coefficient  of the Doppler kernel of the line.
$V^{\scriptscriptstyle (1)}_{\scriptscriptstyle\rm LSR}$ and
$V^{\scriptscriptstyle (2)}_{\scriptscriptstyle\rm LSR}$
are the radial velocities calculated  
using the measured laboratory frequencies $f_{lab}$ and
the predicted theoretical frequencies $f_{cal}$,
respectively (see Table~\ref{T1}).
Given in parentheses are the $1\sigma$ errors in the last digits.
The line width errors are 15-20\%, and the asymmetry coefficient errors are less than 1\%.
}
\begin{tabular}{c r@.l  r@.l  r@.l r@.l r@.l}
\hline\\[-9pt]
No. &
\multicolumn{2}{c}{$f_{obs}$, } &
\multicolumn{2}{c}{\textrm{$FWHM$,} } &
\multicolumn{2}{c}{\textit{As}} &
\multicolumn{2}{c}
{\textrm{$V^{\scriptscriptstyle (1)}_{\scriptscriptstyle\rm LSR}$,}} &
\multicolumn{2}{c}
{\textrm{$V^{\scriptscriptstyle (2)}_{\scriptscriptstyle\rm LSR}$,}} \\
 &
\multicolumn{2}{c}{\textrm{MHz}}  &
\multicolumn{2}{c}{\textrm{ km~s$^{-1}$ } } & 
\multicolumn{2}{c}{\textrm{ } }  & 
\multicolumn{2}{c}{\textrm{ km~s$^{-1}$ } }   &
\multicolumn{2}{c}{\textrm{ km~s$^{-1}$ } } \\
\hline\\[-7pt]
\multicolumn{11}{c}{\textit{ Sgr\,B2(N)}, \textit{1st subband 80.0-89.9 GHz}}\\
1 & 80976&030(10) & 10&0 & $0$&06  & 63&41(19) & 63&72(6) \\
2 & 84405&922(10) & 9&0  &  $0$&02  & 63&52(18) & 63&38(6) \\
3 & 86597&175(9)   & 9&0  & $0$&05   & 63&77(4)   & 63&68(5) \\
4 & 86884&478(9)   & 9&0  & $0$&05   & 63&72(4)   & 63&61(4) \\
5 & 88576&056(10) & 8&0   & $0$&03  & 63&97(17)  & 63&39(5) \\
6 & 88921&240(9)   & 8&0   & $0$&04  & 63&54(17)  & 63&14(5) \\[4pt]

\multicolumn{11}{c}{\textit{ Sgr\,B2(N)}, \textit{2nd subband 89.9-100.2 GHz}}\\
7 & 94521&609(7) & 9&0 & 0&04 & 63&98(16) & 63&98(5) \\
9 & 97561&888(4) & 15&0 & 0&008 & 64&26(2) & 64&239(14) \\[4pt]

\multicolumn{11}{c}{\textit{Sgr\,B2(N)}, \textit{3rd subband 100.2-114.6 GHz} }\\
10 & 100617&577(7) & 9&0 &  $0$&05 & 63&52(15)  & 63&44(5) \\
11 & 111602&917(12) & 8&0 & 0&05 & 63&42(14) &  63&37(5) \\[6pt]

\multicolumn{11}{c}{\textit{Sgr\,B2(M)}, \textit{2nd subband 89.9-100.2 GHz} }\\
8 & 95150&073(11) & 14&0 & 0&01 & 61&08(5) & 60&85(5) \\
9 & 97562&772(14) & 15&0 & 0&03 & 61&54(5) & 61&52(4) \\
\hline
\end{tabular}
\end{table*}

\begin{table*}[h!]
\centering
\label{T3}
\caption{Results of $\Delta\mu/\mu$ estimates from independent groups
of methanol lines. The compared lines are numbered in column 1 according to
Table~\ref{T1}.
The differences between sensitivity coefficients 
$\Delta Q_{\scriptscriptstyle j-i} = Q_{\scriptscriptstyle j} - 
Q_{\scriptscriptstyle i}$ are given in column 2.
The velocity differences
 $\Delta V^{\scriptscriptstyle (k)}_{\scriptscriptstyle i-j} = 
V^{\scriptscriptstyle (k)}_{\scriptscriptstyle i} - 
V^{\scriptscriptstyle (k)}_{\scriptscriptstyle j}$ 
for two types of frequencies~-- measured in laboratory
($k = 1$) and  theoretically calculated ($k = 2$)~--
are listed in columns 3 and 4, respectively.
The corresponding \dmm\ values are presented in the last two
columns 5 and 6.
Given in parentheses are the $1\sigma$ errors in the last digits.
}
\begin{tabular}{c c r@.l r@.l c c}
\hline\\[-7pt]
No.($i-j$) & $\Delta Q_{\scriptscriptstyle j-i}$  &
\multicolumn{2}{c}{$\Delta V^{\scriptscriptstyle (1)}_{\scriptscriptstyle i-j},$}  &
\multicolumn{2}{c}{$\Delta V^{\scriptscriptstyle (2)}_{\scriptscriptstyle i-j},$}  &
$(\Delta\mu/\mu)^{\scriptscriptstyle (1)},$  &
\multicolumn{1}{c}{$(\Delta\mu/\mu)^{\scriptscriptstyle (2)},$}   \\
 &  & 
\multicolumn{2}{c}{km~s$^{-1}$} &  
\multicolumn{2}{c}{km~s$^{-1}$} & $ [\times10^{-7}]$ &  
\multicolumn{1}{c}{$ [\times10^{-7}]$} \\
\hline\\[-7pt]

\multicolumn{8}{c}{\textit{ Sgr\,B2(N)}, \textit{1st subband 80.0--89.9 GHz}  }\\
3--1 & $-3.3$ &0&36(19) & $-0$&04(19) & $-3.6\pm2.3$ & $+0.4\pm2.3$ \\
3--2 & $-1.5$ &0&25(18) & 0&30(18)      & $-5.6\pm4.7$ & $-6.7\pm4.7$ \\
3--5 & $-5.0$ &0&38(18)$^\ast$ & 0&29(18) & $-2.5\pm1.4$ & $-1.9\pm1.4$ \\
3--6 & $-5.0$ & 0&23(18) & 0&54(18) & $-1.5\pm1.3$ & $-3.6\pm1.3$ \\
4--1 & $-3.3$ & 0&31(19) & $-0$&11(19) & $-3.1\pm2.3$ & $+1.1\pm2.3$ \\ 
4--2 & $-1.5$ & 0&20(18) & 0&23(18) & $-4.4\pm4.7$ & $-5.1\pm4.7$ \\
4--5 & $-5.0$ & 0&33(18)$^\ast$  & 0&22(18)  & $-2.2\pm1.3$  & $-1.5\pm1.3$ \\
4--6 & $-5.0$ & 0&18(18)  & 0&47(18) & $-1.2\pm1.3$ & $-3.1\pm1.3$ \\[4pt]

\multicolumn{8}{c}{\textit{Sgr\,B2(N)}, \textit{3rd subband 100.2--114.6 GHz} }\\
10--11 & $-2.7$ & 0&10(21) & 0&07(21) & $-1.2\pm2.5$ & $-0.9\pm2.5$ \\

\multicolumn{6}{r}{\textit{weighted mean} $\langle \Delta\mu/\mu \rangle$:} &
$-2.1\pm0.5$ & $-2.1\pm0.6$ \\[8pt]

\multicolumn{8}{c}{\textit{Sgr\,B2(M)}, \textit{2nd subband 89.9--100.2 GHz}  }\\
8--9 & $+2.9$ & $-0$&46(7) & $-0$&67(7) & $-5.3\pm0.8$ & $-7.7\pm0.8$ \\
\hline\\[-10pt]
\multicolumn{8}{l}{\footnotesize $^\ast$The velocity 
$V^{\scriptscriptstyle (1)}_{\scriptscriptstyle 5}$
was estimated from the
theoretically calculated frequency $f_{cal}$. }
\end{tabular}
\end{table*}

\section{Line selection and processing}
\label{Sec3}

From the published spectra of Sgr\,B2(N) and (M) were selected
($i$) groups of methanol emission lines that are close in frequency and fall within the same frequency range,   
($ii$) strong lines with a signal-to-noise ratio $SNR >60$, ($iii$) with known sensitivity coefficients $Q$ to small changes in  $\mu$, and ($iv$) without significant blending 
the central peaks by adjacent lines of other molecules.
The characteristics of the selected methanol lines are listed
in Table~\ref{T1}, where they are grouped according to the frequency subbands mentioned above:
the second column contains the quantum numbers $J$ and $K$ for the upper $u$ and lower $\ell$ levels, in the third column are the energies of the lower levels $E_\ell$, in the fourth column are the measured laboratory frequencies $f_{lab}$ with errors, in the fifth column~-- theoretically calculated frequencies $f_{cal}$\cite{Xu2008}, and in the last column~-- the sensitivity coefficients $Q$ 
 taken either from \cite{J11} (where $Q = -K_{\mu}$), or calculated in this paper using our procedure described in \cite{LKR}.

Given the complex spatial and dynamical structure of molecular clouds 
discussed in Sect.~\ref{Sec2}, it is first necessary to define what is the 
center of the emission line which probes the gas distribution and kinematics
on spatial scales covered by the aperture of the telescope.
By the center of the emission line we will mean the point at which the
envelope of the line profile reaches its maximum. In the case of a single line 
with symmetric Doppler kernel, this definition coincides exactly with the line
centroid. But in the case of more complex profiles with blends in the wings, this
definition provides more robust estimates of radial velocities compared to the 
line centroid because the Sgr\,B2(N) and (M) spectra contain many weak
emission features, some of which overlap with strong lines of CH$_3$OH.
In this approach, the envelope curve can be determined by a
multicomponent Gaussian model  (with the smallest number of Gaussian
subcomponents) and its fitting to the observed methanol line under the 
condition of achieving the global minimum of the objective function  at 
$\chi^2_\nu  \la 1$, where $\nu$ is the number of degrees of freedom.

Such fits are displayed in Fig.~\ref{F1} where the original spectra are
presented by black dots and the resulting envelope curves are shown in red.
The horizontal green line indicates the range in which 
the global minimum of the $\chi^2$ function was calculated.
For each line the $SNR$ and $\chi^2$ (per degree of freedom)
are labeled at the top left hand corner, while the object name and the line
laboratory frequency are displayed at the top right hand corner. The identified
neighboring blends plotted in Figs.~2 and 5 in \cite{Bel} are marked by vertical
blue bars in our Fig.~\ref{F1} as well. The observed red wing asymmetry of
all Sgr\,B2(N) lines is caused by a weaker methanol emission from the
second hot core N2 shifted at $\approx 11$ \kms. These secondary velocity
components will not be analyzed , because they do not provide sufficient
accuracy with respect to line parameters.

The calculated fitting parameters of the envelope curves~-
the observed frequency of the line peak ($f_{obs}$) 
and the line width ($FWHM$)
for each transition are listed in Table ~\ref{T2} in the second and third columns, respectively. The fourth column shows the asymmetry coefficients $As$ of the Doppler kernel of the line (for a purely Gaussian profile, $As = 0$).
The fifth and sixth columns give
the radial velocities $V_{\scriptscriptstyle\rm LSR}$, calculated following 
the radio astronomical convention: 
\begin{equation}
V^{\scriptscriptstyle (1)}_{\scriptscriptstyle \rm LSR} = c\left( 1 - {f_{obs}}/{f_{lab}}\right),\,\,\,
V^{\scriptscriptstyle (2)}_{\scriptscriptstyle \rm LSR} = c\left( 1 - {f_{obs}}/{f_{cal}}\right).
\label{E3}
\end{equation}

The measured radial velocity errors 
$\sigma_{\scriptscriptstyle V_{\scriptscriptstyle\rm LSR}}$ 
include both errors in measured line positions
$\sigma_{f_{obs}}$
and uncertainties in laboratory frequencies
$\sigma_{f_{lab}}$,
which are indicated in Table~\ref{T1}:
 \begin{equation}
\sigma_{\scriptscriptstyle V_{\scriptscriptstyle\rm LSR}} = c\frac{f_{obs}}{f_{lab}}\sqrt{\delta^2_{obs} + \delta^2_{lab}}\, ,
\label{E56}
\end{equation}
where $\delta_{obs} = \sigma_{f_{obs}}/f_{obs}$ and
$\delta_{lab} = \sigma_{f_{lab}}/f_{lab}$.

A comparison of the measured frequency errors from Table~\ref{T2} with the
laboratory frequency errors from Table~\ref{T1} shows that the radial velocity
errors are dominated by the laboratory errors.
The theoretical errors were calculated using the least squares method, in which the observed laboratory frequencies and their uncertainties were used to adjust the parameters of the Hamiltonian, which was used to calculate the frequencies $f_{cal}$\cite{Xu2008}.
Therefore, in further calculations of $\Delta\mu/\mu$, errors of real laboratory measurements were used, rather than errors of calculated frequencies.
These errors in most cases amount to $\sim 10$\% of the channel width,  and
the change in radial velocity $\sim 1$~\ms\ due to the rotation of the Earth between the receiver's heterodyne resets can be ignored.

We also did not take into account systematic errors due to the non-linearity of converting the frequency scale
to the velocity scale, which can reach several kHz for objects from the Galactic disk when using the standard
CLASS\cite{IRAM} software package.
In our analysis, these errors are insignificant, since we processed the line profiles in a frequency scale
that is linear for the  autocorrelators \cite{Tif}, and only then converted the resulting line centers to
radial velocities according to the recommendations from \cite{Tif}.

It should be noted that there is another source of systematic errors related
to inaccuracies in pointing the telescope at a series of observations and then adding up individual exposures to increase
the signal-to-noise ratio.
Pointing errors ($2^{\prime\prime}-3^{\prime\prime}$) can lead to additional
uncertainties in the positions of the lines, since the telescope beam may not always cover exactly the same area in the molecular cloud.
It is not possible to assess such errors in individual exposures.
However, the procedure
of adding up a large number of independent exposures should 
reduce such uncertainties in the integral spectra, and comparing
the measurement results with data obtained with another telescope \cite{VL25}
can serve as an additional control for possible systematic shifts.

The line widths shown in Table ~\ref{T2}, in addition to the kinematic characteristics of the radiated area, also reflect the uncertainties
of telescope pointing between individual exposures.
Errors in the line widths and errors in the measured
fluxes in emission lines reach values of 15-20\% at 3~mm \cite{Bel}.
Therefore, the given line widths can only serve as an additional filter
when selecting line pairs for estimates of $\Delta\mu/\mu$, whose widths differ by no more than 1--2 \kms.

This condition is satisfied by lines from the 1st and 3rd subbands in the Sgr\,B2(N) spectrum and lines from the 2nd subband in the Sgr\,B2(M) spectrum.
Line 9 from the 2nd subband in the Sgr\,B2(N) spectrum is  essentially widened
due to a significant contribution from the N2 component, whose effect on the width of the weaker line 7 from the same subband is less noticeable 
(see Fig.~\ref{F1}). Therefore, lines 7 and 9 are unsuitable for precision estimates of \dmm, despite the fact that they
have the highest $SNR$ among the selected methanol lines in the Sgr\,B2(N) spectrum. The remaining lines listed in Table~\ref{T2} have small values of the asymmetry coefficients  ($As \ll 0.1$) and approximately the same widths of the Doppler kernels, which indicates their origin from the same spatial region.

\begin{figure}[h!]
\centering
\includegraphics[width=0.8\textwidth]{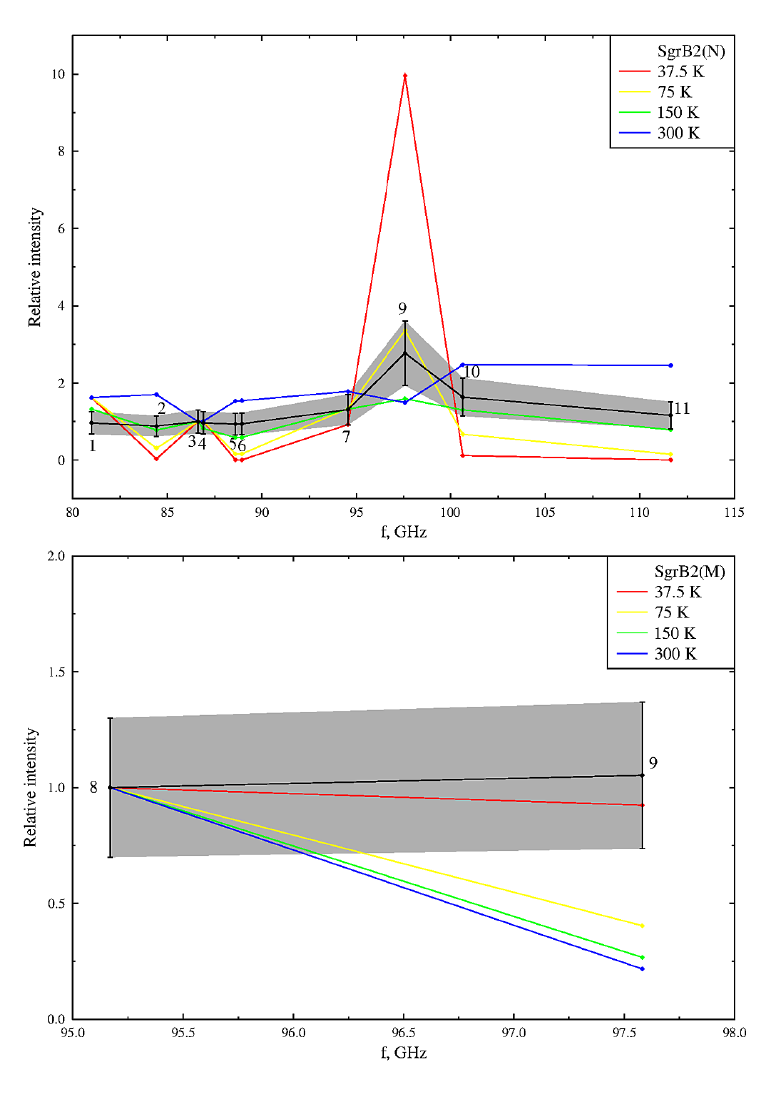}
\caption{\small Comparison of peak relative intensities of observed lines
(black) with theoretical ones (colored lines).
The numbers correspond to the line numbers from Table~\ref{T1}.
The upper panel shows the relative line intensities in Sgr\,B2(N), normalized
to the 86.615 GHz line (No.3), and on the lower panel~-- for Sgr\,B2(M)~-- relative to the 95.169 GHz line (No.8).
The shaded areas show the confidence zones $\pm 1\sigma$, calculated with a maximum flux calibration error of 20\% \cite{Bel}. 
}
\label{F2}
\end{figure}

In addition to controlling the shape of the line kernels, it is possible to compare peak values of relative intensities with theoretical values calculated for certain excitation temperatures. For this purpose, we used the procedure from CDMS \cite{cdms}. Figure~\ref{F2} shows the results of such a comparison~-- the colored lines display the theoretical relative intensities for different
excitation  temperatures in the optically thin case: 
$T_{ex} = 300$~K, 150~K, 75~K and 37.5~K, and in black~--
 the observed relative intensities normalized to the
intensity of the 86.615 GHz line (No.3 on the upper panel, Sgr\,B2(N)) and on the 95.169 GHz line intensity (No. 8 on the lower panel, Sgr\,B2(M)).
The shaded areas show the confidence zones $\pm 1\sigma$, calculated with a maximum flux calibration error of 20\% \cite{Bel}.
The numbers on the panels correspond to the line numbers from 
Table~\ref{T1}.

From Fig.~\ref{F2} it can be seen that the relative intensities of the lines selected from the spectrum of Sgr\,B2(M)
correspond to a low excitation temperature of 37.5 K,
while in Sgr\,B2(N) the observed pattern is more complicated.
The relative intensities of the lines from the 1st subband (80--89.9 GHz) correspond
to a higher excitation temperature of 150 K, as well as lines from the 3rd subband (100.2--114.6 GHz)~-- all these lines probably originate from the same area~--
 as evidenced by their similar profiles and velocities.
However, line 9 from the 2nd subband (89.9--100.2 GHz) is most suitable for an
excitation  temperature of about 75 K.
At the same time, the increased relative intensity at point 9 compared to the theoretically
expected value for $T_{\rm ex} = 150$~K (green line in the figure) may also indicate a significant contribution of the internal
blend from the N2 source, as discussed above.
And this is another reason not to consider lines 7 and 9
from the 2nd subband  in the calculations \dmm\ for Sgr\,B2(N).

\section{$\Delta\mu/\mu$ estimates}
\label{Sec4}

Differential measurements of the fundamental physical constant $\mu$ 
are made from
comparing pairs of methanol lines with known sensitivity coefficients $Q_i$, $Q_j$, which are formed in the same region
of the molecular cloud and have 
radial velocities $V_i^{(k)}$, $V_j^{(k)}$ \cite{LKR}:
 \begin{equation}
(\Delta\mu/\mu)^{(k)} \equiv (\mu_{\rm obs} - \mu_{\rm lab})/\mu_{\rm lab} =
(V_i^{(k)} - V_j^{(k)})/[c(Q_j - Q_i)]\, ,
\label{E3}
\end{equation}
where $c$ is the speed of light, $k = 1$ or 2 for radial velocities
calculated from laboratory measured frequencies or theoretical ones, respectively.

There is some freedom of choice when calculating the radial velocity difference
between the pairs of methanol lines that
are used to estimate \dmm.
Namely, the most sensitive estimates are obtained with the largest modulo denominator,
$\Delta Q_{j-i} = Q_j - Q_i$, in equation (\ref{E3}).
Based on these considerations, the \dmm\ values presented in Table~3
for two sets of reference frequencies $f_{lab}$ and $f_{cal}$ were calculated.
Using two independent measurements of the position of lines 3 and 4 having the same values of $Q$,
the largest for lines from the 1st frequency subband,
it is possible to make  8 pairs of differences $\Delta V_{i-j}$ and 8 corresponding values of
\dmm\ specified in  the fifth and sixth columns of Table~3.
The total sample $(\Delta\mu/\mu)_1, (\Delta\mu/\mu)_2, \ldots, (\Delta\mu/\mu)_n$ for Sgr\,B2(N)
also includes the measurement of one pair of lines from the 3rd subband,
i.e. , the sample size is $n = 9$.
Meanwhile it is necessary to take into account the following circumstance.

The measurement errors $\sigma_{\Delta\mu/\mu}$ are not statistically independent
because the differences $\Delta V_{i-j}$
 were calculated with respect to two lines  3 and 4 in the 1st subband Sgr\,B2(N).
This leads to correlated  \dmm\ values (for detail, see \cite{L2010, VKL}).
Such correlations should increase the calculated errors $\sigma_{\Delta\mu/\mu}$
by a correction factor
\begin{equation}
\gamma = (1-r^2)^{-1/2}\, ,
\label{E57}
\end{equation}
where $r$~is the mean correlation coefficient, which can be estimated
from equation (16) in \cite{VKL}.

The value of $r$ equals 0.5 for 8 pairs of lines from the sample under consideration, which corresponds to the correction factor
$\gamma = 1.15$:
\begin{equation}
\sigma^{\scriptscriptstyle \rm cor}_{\Delta\mu/\mu} =\gamma \sigma_{\Delta\mu/\mu}\, .
\label{E58}
\end{equation}
One pair of lines 10--11 from the 3rd subband in the Sgr\,B2(N) spectrum
gives \dmm, which is included in the sample with its own
error $\sigma_{\Delta\mu/\mu} = 2.5\times 10^{-7}$.

The \dmm\ estimate in Sgr\,B2(M) is calculated using a pair of lines 8 and 9: 
$\Delta\mu/\mu = (-5.3\pm 0.8) \times 10^{-7}$.
Since it was not possible to select more methanol lines in the Sgr\,B2(M) spectrum that meet the selection criteria, the value of \dmm\ in this case cannot be considered statistically significant, despite the small error 
$\sigma_{\Delta\mu/\mu} = 0.8 \times 10^{-7}$.
Therefore, our main result will be based on the sample Sgr\,B2(N), consisting
of $n=9$ unequal accuracy data of $\Delta\mu/\mu$.
The  processing of unequal-accuracy measurements is described in detail in \cite{A72}.

Note that three of the selected lines~-- 1, 6, and especially 5~-- have very different values ​​of $f_{lab}$ and $f_{cal}$ (see Table~\ref{T1}).
Therefore, the velocity of line 5 $V^{\scriptscriptstyle (1)}_{\scriptscriptstyle 5}$ was determined from the theoretically calculated frequency $f_{cal}$, since the measured laboratory frequency of this line gives  $V^{\scriptscriptstyle (1)}_{\scriptscriptstyle 5} = 63.97$ \kms, which differs significantly from the velocity of line 6 $V^{\scriptscriptstyle (1)}_{\scriptscriptstyle 6} = 63.54$ \kms, which has the same sensitivity coefficient.
At the same time, lines 3 and 4, which also have the same $Q$ values ​​and for which laboratory frequencies are known with good accuracy (5 kHz), give approximately the same velocities (see Table~\ref{T2}).

Moreover, as noted in \cite{Mull}, theoretically calculated frequencies turn out to be closer to laboratory ones in those cases when the latter are measured with an accuracy higher than 50 kHz. Therefore, we present a second sample of \dmm\ values ​​calculated using the frequencies $f_{cal}$.
Both samples are presented in Table~3 in the fifth and sixth columns, respectively.

Following the algorithm for processing series of unequal-accuracy measurements \cite{A72}, we obtain the following weighted mean values:
\begin{equation}
\langle \Delta\mu/\mu \rangle^{(1)}  = (-2.1\pm0.5)\times10^{-7}\, ,
\label{E59}
\end{equation}
and
\begin{equation}
\langle \Delta\mu/\mu \rangle^{(2)}  = (-2.1\pm0.6)\times10^{-7}.
\label{E60}
\end{equation}

The parameter $\kappa$, which characterizes the presence of significant systematic errors in the sample
for a value of $\kappa>2$ (ratio 4.138 in \cite{A72}), is equal to 1.42 for the first sample Sgr\,B2(N) and 0.26 for the second sample,
i.e., no obvious systematic errors, random outliers, or underestimates of the errors
of the measured values ​​are detected in these samples.

Comparison of samples
$(\Delta\mu/\mu)^{\scriptscriptstyle (1)}$ and 
$(\Delta\mu/\mu)^{\scriptscriptstyle (2)}$ in Table~3 shows a high dependence of the \dmm\ estimates  on the reference frequencies $f_{lab}$ and $f_{cal}$.
Nevertheless, the weighted mean values ​​$\langle \Delta\mu/\mu \rangle^{\scriptscriptstyle (1)} = \langle \Delta\mu/\mu \rangle^{\scriptscriptstyle (2)} $ within the $1\sigma$ uncertainty intervals.
Therefore, replacing the laboratory frequency of the 5th line with the theoretically calculated one does not lead to fictitious estimates of \dmm.

For the final result, we take the weighted mean value 
$\langle \Delta\mu/\mu \rangle^{\scriptscriptstyle (2)} = (-2.1 \pm 0.6) \times 10^{-7}$, obtained from the measured frequencies $f_{obs}$ of the methanol line centers in Sgr\,B2(N) and the theoretically calculated frequencies $f_{cal}$.

\section{Conclusions}
\label{Sec5}

In conclusion, we note that,
both the previous estimate
of \dmm\ in the high-frequency spectrum of Sgr\,B2(N) at 0.6~mm, obtained with the {\it Herschel} space telescope,
and the estimates at 3~mm in the spectra of Sgr\,B2(N) and (M), obtained with the IRAM 30-m telescope,
lead to negative \dmm\ values ​​at the level of a few$\times10^{-7}$.

This result may indicate  possible spatial variations of  $\mu$ in the direction of the  molecular cloud Sgr\,B2.
However, the absolute values ​​of \dmm\ cannot yet be reconciled within the  uncertainty interval $\pm 1\sigma$.
Apparently, one of the reasons is the insufficiently high spectral resolution of the spectra used.

Second, it would be desirable to have more accurate laboratory measurements of the methanol line frequencies to reduce errors in the measured radial velocities and to verify frequency shifts between the CH$_3$OH lines analyzed in this study.

Finally, a search for other molecules in the vicinity of the Galactic center suitable for estimating \dmm\ is highly desirable for independent future tests of the obtained result.
It would be preferable to also carry out such tests on interferometers with high spatial resolution.

\subsection*{Funding}
This work was supported by the Ministry of Science and Higher Education
of the Russian Federation, state assignment No.~FFUG-2024-0002
for the Ioffe Institute.
No additional grants to carry out or direct this particular research were obtained.

\subsection*{Conflict of interest} 
The authors of this work declare that they have no conflicts of interest.

\subsection*{Acknowledgments} 
The authors would like to thank Arnaud Belloche for additional information on the observations and calibration of the spectra of Sgr\,B2(N) and (M) obtained with the IRAM 30-m telescope.
We also thank our anonymous reviewer for constructive comments.


\begin{thebibliography}{99}

\bibitem{Dir37} P. A. M. Dirac, Nature {\bf 139}, 323 (1937).

\bibitem{Kh04} J. Khoury, \& A. Weltman, Phys. Rev. Lett. {\bf 93}, 171104 (2004).

\bibitem{Flam08} V. V. Flambaum and E. V. Shuryak, in Nuclei and Mesoscopic Physics, eds.
P. Danielewicz,  P. Piecuch, \& V. Zelevinsky, AIP Conf. Series, vol. 995, pp.1-11, AIP Publishing, New York (2008). 

\bibitem{Ub18} W. Ubachs, Space Sci. Rev. {\bf 214}, 3 (2018).

\bibitem{Bar02} J. Barrow, H. Sandvik, \& J. Magueijo, Phys. Rev. D {\bf 65}, 063504 (2002).

\bibitem{Uz24} J.-P. Uzan, Living Rev. Relativity {\bf 26}, 6 (2025).

\bibitem{Kan} N. Kanekar, W. Ubachs,  K. M. Menten,   J. Bagdonaite,  A. Brunthaler, 
C. Henkel, S. Muller, H. L. Bethlem, \&  M. Dapr\`a,  
Mon. Not. R. Astron. Soc. {\bf 448}, L104 (2015). 

\bibitem{L11} S. A. Levshakov, M. G. Kozlov, \& D. Reimers,
Astrophys. J. {\bf 738}, 26 (2011).

\bibitem{D17} M. Dapr\`a, C. Henkel, S. A. Levshakov, K. M. Menten, S. Muller, H. L. Bethlem,  S. Leurini, A. V. Lapinov, \& W. Ubachs,  
Mon. Not. R. Astron. Soc. {\bf 472}, 4434 (2017).

\bibitem{VL24}  J. S. Vorotyntseva, \&  S. A. Levshakov,
JETP Letters {\bf 119}, 649 (2024).

\bibitem{A19} A. Amorim, M. Baub\"ock, J. P. Berger et al., Phys. Rev. Lett. {\bf 122}, 101102 (2019).

\bibitem{Hees20} A. Hees, T. Do, B. M. Roberts et al., Phys. Rev. Lett. {\bf 124}, 081101 (2020).

\bibitem{VL25} J. S. Vorotyntseva, \&  S. A. Levshakov,
JETP Letters {\bf 121}, 589  (2025).

\bibitem{RMZ}   M. J. Reid,  K. M. Menten,  X. W. Zheng, A. Brunthaler,  \&  Y. A. Xu, 
Astrophys. J. {\bf 705}, 1548 (2009).

\bibitem{Schw} A. Schw\"orer, \'A. S\'anchez-Monge, P. Schilke et al., Astron. Astrophys.  {\bf 628}, A46 (2019).

\bibitem{Sc}  N. Z. Scoville,  P. M. Solomon,  \& A. A. Penzias, Astrophys. J.
{\bf 201}, 352 (1975).

\bibitem{Bel} A. Belloche,  H. S. P. M\"uller,  K. M. Menten, P. Schilke, \& C. Comito, 
Astron. Astrophys.  {\bf 559}, A47 (2013).


\bibitem{Bel13http} http://cdsarc.u-strasburg.fr/viz-bin/qcat?J/A+A/559/A47


\bibitem{CLASS} http://www.iram.fr/IRAMFR/GILDAS 


\bibitem{Xu1997} L.-H. Xu, \& F. J. Lovas,  J. Phys. Chem. Ref. Data {\bf 26}, 1 (1997). 

\bibitem{Mull}  H. S. P. M\"uller, K. M.  Menten, \& H. M\"ader, 
Astron. Astrophys. {\bf 428}, 1019 (2004).

\bibitem{Xu2008} L.-H. Xu, J. Fisher, R. M. Lees, H. Y. Shi, J. T. Hougen, J. C. Pearson,
B. J. Drouin, G. A. Blake, \& R. Braakman, J. Mol. Spec. {\bf 251}, 305 (2008). 

\bibitem{J11}  P. Jansen, L. H. Xu, I. Kleiner, W. Ubachs, \& H. L. Bethlem, 
 Phys. Rev. Lett., {\bf 106}, 100801 (2011).

\bibitem{LKR}  S. A. Levshakov, M. G. Kozlov,  \& D. Reimers, 
Astrophys. J. {\bf 738}, 26 (2011).

\bibitem{IRAM} J. Pety, \& S. Bardeau, IRAM Memo 2011-4 (2019).

\bibitem{Tif} W. G. Tifft, \& W. K. Huchtmeier, Astron. Astrophys. Suppl. Ser. {\bf 84}, 47 (1990).

\bibitem{cdms} Cologne Database for Molecular Spectroscopy, https://cdms.astro.uni-koeln.de/cgi-bin/cdmssearch

\bibitem{L2010} S. A. Levshakov, P. Molaro, A. V. Lapinov, D. Reimers, C. Henkel, \& T. Sakai, 
Astron. Astrophys. {\bf 512}, A44 (2010). 

\bibitem{VKL} J. S. Vorotyntseva, M. G. Kozlov, \& S. A. Levshakov, 
Mon. Not. R. Astron. Soc. {\bf 527}, 2750 (2024).

\bibitem{A72} T. A. Agekyan, {\it The basis of the theory of errors for astronomers and physicists} (Nauka, Moscow, 1972), p.144.

\end{thebibliography}
\end{document}